\documentclass[11pt]{article}

\usepackage[margin=1.4in]{geometry}

\usepackage[labelfont=bf]{caption} % Bold figure number
\usepackage[font={small}]{caption}   %% FIGURE CAPTION FONT MODIFICATION

\usepackage{graphicx}% Include figure files
\usepackage{dcolumn}% Align table columns on decimal point
\usepackage{bm}% bold math
\usepackage{hyperref}
\usepackage[utf8]{inputenc}
\usepackage{lipsum}  
\usepackage{chngpage}
\usepackage{amsmath}
\usepackage{mathtools}
\usepackage{authblk}
\usepackage{xfrac}
\usepackage{placeins}
\usepackage{verbatim}

\usepackage[numbers, sort&compress]{natbib}  %%% For compressing the reference numbering

\usepackage[usenames,dvipsnames]{color}    % ADDED FOR COLOR NAMES
\usepackage{soul}   % SOUL Added for underlines
\setstcolor{red} % strikethrough color
\setul{}{1.5pt}  % line thickness

\begin{document}

\title{Synthetic phonons enable nonreciprocal coupling \\to arbitrary resonator networks}

\author[1]{Christopher W Peterson}
\author[2]{Seunghwi Kim}
\author[1]{Jennifer T Bernhard}
\author[2,$\ast$]{\mbox{Gaurav Bahl}}

\affil[ ]{\footnotesize{University of Illinois at Urbana-Champaign}}
\affil[1]{Department of Electrical and Computer Engineering}
\affil[2]{Department of Mechanical Science and Engineering}
\affil[$\ast$]{To whom correspondence should be addressed; bahl@illinois.edu}

\date{\today}

\maketitle

\begin{abstract}
Inducing nonreciprocal wave propagation is a fundamental challenge across a wide range of physical systems in electromagnetics, optics, and acoustics.
Recent efforts to create nonreciprocal devices have departed from established magneto-optic methods and instead exploited momentum based techniques such as coherent spatiotemporal modulation of resonators and waveguides.
However, to date the nonreciprocal frequency responses that such devices can achieve have been limited, mainly to either broadband or Lorentzian-shaped transfer functions.
Here we show that nonreciprocal coupling between waveguides and resonator networks enables the creation of devices with customizable nonreciprocal frequency responses.
We create nonreciprocal coupling through the action of synthetic phonons, which emulate propagating phonons and can scatter light between guided and resonant modes that differ in both frequency and momentum.
We implement nonreciprocal coupling in microstrip circuits and experimentally demonstrate both elementary nonreciprocal functions such as isolation and gyration as well as reconfigurable, higher-order nonreciprocal filters.
Our results suggest nonreciprocal coupling as platform for a broad class of customizable nonreciprocal systems, adaptable to all wave phenomena.
\end{abstract}

%
%%%%%%%%%%%%%%%%%%%%%%%%%%%%%%%
%%%%%%%%%%%%%%%%%%%%%%%%%%%%%%%
\section*{Introduction}
%%%%%%%%%%%%%%%%%%%%%%%%%%%%%%%
%%%%%%%%%%%%%%%%%%%%%%%%%%%%%%%
%

\begin{figure}
\begin{adjustwidth}{-.4in}{-.4in}
    \centering
    \makebox[\textwidth][c]{\includegraphics[width = \linewidth]{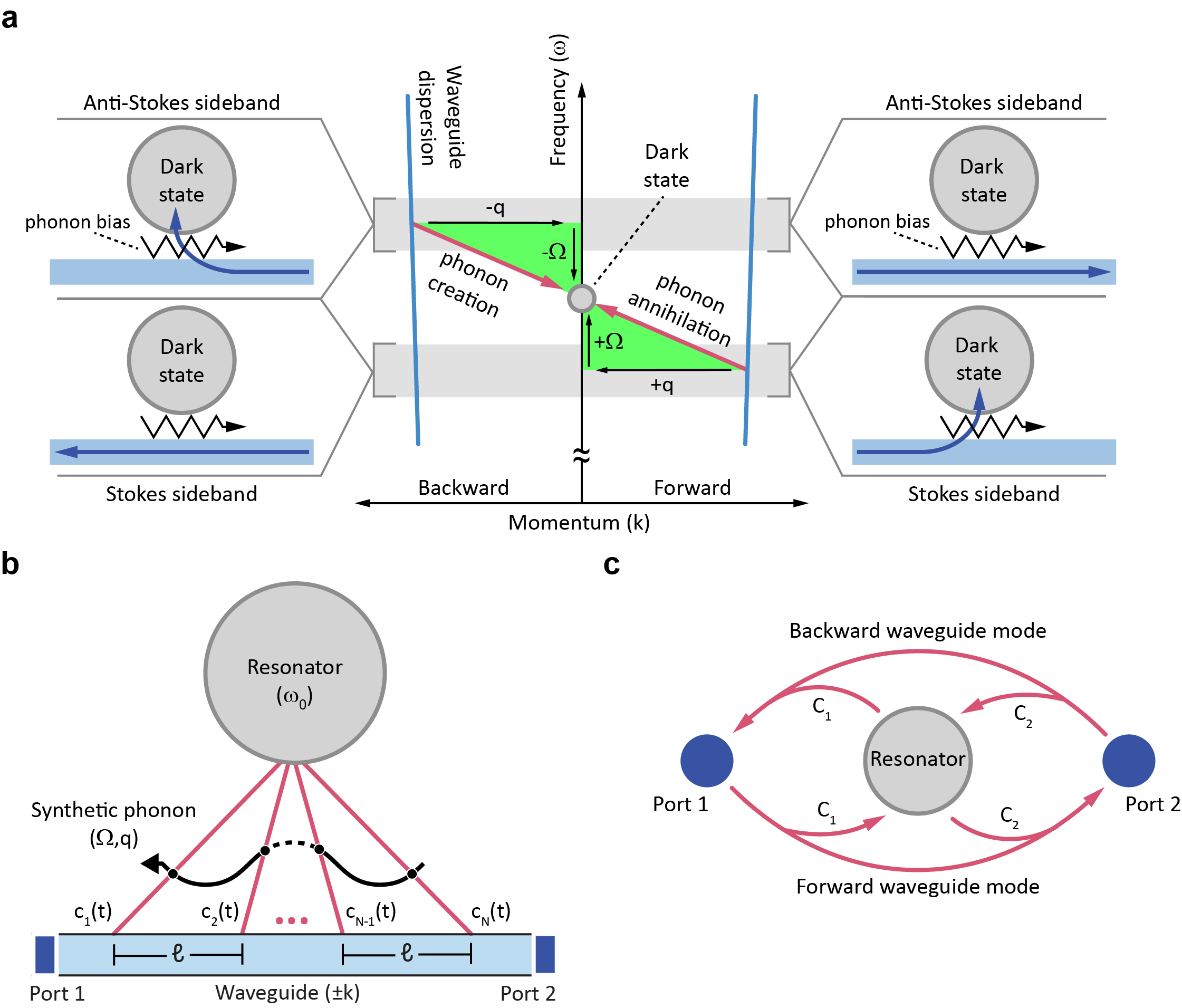}}%
    \caption{Theoretical description of nonreciprocal coupling.
    (a) Interactions between photons and phonons enable coupling to photonic dark states by modifying the phase matching condition. Phonon-enabled coupling occurs only when the frequency and momentum difference between the waveguide mode and resonator mode are matched by the phonon frequency and momentum. This coupling is inherently nonreciprocal: the forward (right-traveling) guided mode only couples to the dark resonant state through phonon annihilation at the Stokes sideband frequency ($\omega_0 - \Omega$), and the backward (left-traveling) guided mode only couples to the resonator through phonon creation at the anti-Stokes sideband frequency ($\omega_0 + \Omega$).
    (b) A resonator and two-port waveguide are coupled at multiple spatially separated sites. The coupling at these sites can be modulated to create synthetic phonons, which can enable nonreciprocal coupling between the waveguide and resonator. 
    (c) Schematic describing the coupling constants $C_1$ and $C_2$. $C_1$ describes coupling into the resonator from the forward waveguide mode and coupling out of the resonator to the backward waveguide mode and is associated with port 1. $C_2$ describes coupling into the resonator from the backward waveguide mode and coupling out of the resonator to the forward waveguide mode and is associated with port 2. 
    }
    \label{fig1}
\end{adjustwidth}
\end{figure}

%
The interactions between elementary particles and quasi-particles (electrons, photons, phonons) are dictated by both momentum and energy conservation; this is broadly termed as phase matching.
These conservation laws are especially important in the field of nonlinear optics \cite{Boyd}, as inelastic scattering of light involving the creation or annihilation of propagating phonons can produce large shifts in photon momentum \cite{Bahl2, SKim, SKim2}.
Phonon-assisted momentum shifts in turn permit unique phenomena such as indirect interband photonic transitions \cite{Hwang, Yu, Bahl, Pant, Kittlaus}, where light is scattered between optical modes that differ in both frequency and momentum.
Indirect interband transitions and similar processes associated with significant momentum shifts have recently been identified as promising tools for inducing nonreciprocal transmission of light \cite{Hwang, Yu, Kamal, Kang, Hafezi, Poulton, Lira, Estep, Qin, Kim, Dong, EstepIEEE, Shen, Kim2, Shii} and sound \cite{Fleury} without reliance on magnetic fields.
Devices based on these effects break Lortenz reciprocity because momentum shifts are not symmetric under time-reversal \cite{Casimir}, e.g. if an incoming wave scatters and gains momentum, the same wave will lose momentum in the time-reversed process.
To date, nonreciprocal devices based on indirect photonic transitions have exclusively relied on scattering between co-propagating modes in waveguides \cite{Yu, Poulton, Kang, Lira} or resonators \cite{Kim, Dong, Kim2}.
In this paper we demonstrate that indirect transitions can also be induced between a guided mode and a stationary resonant mode (Fig. \ref{fig1}a) through the action of synthetic phonons, which emulate propagating phonons but have independently controlled frequency and momentum.
These indirect transitions effectively generate nonreciprocal coupling between the guided and resonant mode, as only one propagation direction is coupled to the resonance.
As we will show, the power of nonreciprocal coupling is that highly tailorable, reconfigurable nonreciprocal transfer functions can be arranged using conventional waveguides and resonators.
It is important to note a major difference between indirect transitions among co-propagating modes and nonreciprocal coupling between a guided and resonant mode.
Co-propagating modes in a waveguide or resonator are orthogonal and thus, by definition, are not coupled unless by an applied bias.
In contrast, resonant modes often couple to guided modes.
This coupling is reciprocal, making it undesirable in systems utilizing nonreciprocal coupling.
To ensure that only nonreciprocal coupling occurs in our system, we intentionally create a phase mismatch between the guided and resonant modes.
Throughout this paper we refer to such phase mismatched resonant modes as dark states because, analogous to atomic dark states \cite{Boller, Arimondo}, they are localized resonances that cannot emit or absorb light. 
Dark states have previously been studied for applications in photonics \cite{OptomechDark, Guo, Yang2} and have several properties, namely suppressed emission and long lifetime, that are especially useful for creating nonreciprocal devices.
We begin our discussion with a generalized system consisting of a waveguide and resonator that are coupled at multiple spatially separated coupling sites, as in Fig. \ref{fig1}b. 
We first describe how the interference between these coupling sites can be exploited to suppress interactions between the waveguide and resonator and create a dark state. 
We then show that time-modulation of the coupling constant at each site emulates the effect of driven phonons and can re-enable interactions between the prepared dark state with only one propagation direction within the waveguide (Fig. \ref{fig1}a).
The form of our proposed coupling rate modulation, which we term a `synthetic phonon', is depicted in Fig. \ref{fig1}b.
Synthetic phonons are analogous to phonons in a crystal lattice: coupling sites are equivalent to lattice sites and the varying coupling rate is equivalent to atomic displacement. 
In contrast to phonons in a crystal, the synthetic phonon momentum is externally controlled and does not depend on an intrinsic dispersion relation, allowing synthetic phonons to be applied in a wide variety of situations without the added complexity of dispersion engineering \cite{Lira}.
We experimentally realize synthetic phonons in microwave frequency microstrip circuits by sinusoidally modulating the capacitive coupling rate of spatially-separated variable capacitors.
Using these synthetic phonons, we experimentally demonstrate several distinct nonreciprocal effects including extreme isolation contrast ($> 82$ dB), nonreciprocal phase shifts, and higher-order nonreciprocal filters.
%

%%%%%%%%%%%%%%%%%%
\section*{Results}
%%%%%%%%%%%%%%%%%
%
\subsection*{Nonreciprocal coupling to engineered dark states}

We consider our representative system (Fig. \ref{fig1}b) consisting of a two-port waveguide having a frequency dependent propagation constant $k$ and a resonator supporting a single mode at angular frequency $\omega_0$.
The resonator is side-coupled to the waveguide at $N$ independent sites that are evenly %periodically
separated on the waveguide by a constant length $\ell$. 
For simplicity, we assume that each coupling site is located at the same spatial location on the resonator, and that the waveguide is lossless and only supports a single mode.
We define forward propagation in the waveguide from port 1 toward port 2. 

This system can be characterized by analyzing the coupling between the waveguide and the resonator using the framework of temporal coupled-mode theory \cite{Haus, CMT}.
Since each coupling site is independent, the coupling constants $C_1$ and $C_2$ between the waveguide and resonator (see Fig. \ref{fig1}c) are evaluated as a superposition of the contributions from each site,
\begin{equation} \label{coupling}
%\begin{split}
    C_1 = \sum_{n=1}^{N} c_n e^{-i k \ell (n-1)}  ~,~ C_2 = \sum_{n=1}^{N} c_n e^{i k \ell (n-1)}  ~, 
%\end{split}
\end{equation}
where $c_n$ is the coupling constant at the $n^{th}$ site.
The exponential term in these definitions accounts for propagation in the waveguide between adjacent coupling sites spaced by $\ell$, and differs between $C_1$ and $C_2$ due to the opposite propagation directions.
The coupling constants are also related to the resonator's decay, which can be described by the decay rate $\gamma = (\kappa_i + \kappa_{ex}) /2 $.
Here, $\kappa_i$ is the intrinsic decay rate of the resonator and $\kappa_{ex} = | C_1 | ^2 + | C_2 | ^2$ is the external decay rate due to coupling with the waveguide \cite{CMT}.
Equation (\ref{coupling}) reveals that the contribution from the $n^{\textnormal{th}}$ coupling site carries a phase $k \ell (n - 1)$. 
When summed, these contributions interfere such that the maximum coupling rate occurs only if all $N$ contributions are in-phase (phase matched coupling). 
The coupling rate decreases away from this maximum and reaches zero when the contributions perfectly destructively interfere.
In the case of a complete phase mismatch, we obtain $\kappa_{ex} = 0$ and the resonator can be classified as a dark state since it cannot be excited by (or decay to) the waveguide.
Since phase matching in this system is determined by the product $k \ell$, it is possible to arrange a dark state from an arbitrary waveguide and resonator by selecting the appropriate $\ell$.
We now discuss how a dark state created by a total phase mismatch can be coupled to the accompanying waveguide through a synthetic phonon bias (Fig.~\ref{fig1}a,b).
We consider synthetic phonons having angular frequency $\Omega$, momentum $q$, and amplitude $\delta_c$, which are a modulation of each site's coupling rate %via the expression
\begin{equation} \label{modulation}
    c_n = c_0 + \delta_c \cos(\Omega t - q \ell (n-1)) ~.
\end{equation}
The product $q \ell$ is equivalent to a phase offset on the modulation applied to adjacent sites, thus any phonon momentum $q$ can be selected by modulating each site with a phase offset $\theta_n = q \ell (n-1)$.
When this spatiotemporally modulated coupling is substituted into Equation (\ref{coupling}) it is instructive to separate the resulting terms into frequency components as follows:
\begin{equation} \label{mod_coupling1}
\begin{split}
   C_1 = \overbrace{ c_0 \sum_{n=1}^{N} e^{-i k \ell (n-1)} }^{C_1^0}  + \overbrace{ \frac{\delta_c}{2} e^{i \Omega t} \sum_{n=1}^{N} e^{- i (k + q) \ell (n-1)} }^{C_1^+} 
    + \overbrace{ \frac{\delta_c}{2} e^{-i \Omega t} \sum_{n=1}^{N} e^{-i (k - q) \ell (n-1)} }^{C_1^-} ~,
\end{split}
\end{equation}
\begin{equation} \label{mod_coupling2}
\begin{split}
   C_2 = \overbrace{ c_0 \sum_{n=1}^{N} e^{i k \ell (n-1)} }^{C_2^0} + \overbrace{ \frac{\delta_c}{2} e^{i \Omega t} \sum_{n=1}^{N} e^{i (k - q) \ell (n-1)} }^{C_2^+} 
   + \overbrace{ \frac{\delta_c}{2} e^{-i \Omega t} \sum_{n=1}^{N} e^{i (k + q) \ell (n-1)} }^{C_2^-} ~.
\end{split}
\end{equation}
For brevity, from here we refer to the terms that make up the coupling constants as $C_{m} = C_{m}^0 + C_{m}^+ + C_{m}^-$ for $m = 1 , 2$.
The first term $C_{m}^0$ does not depend on the modulation amplitude $\delta_c$ and describes coupling which would occur without synthetic phonons. 
The remaining terms only describe coupling enabled by interactions with synthetic phonons: $C_{m}^+$ corresponds to coupling where a synthetic phonon is annihilated and the photon shifts up in frequency, and $C_{m}^-$ corresponds to coupling where a synthetic phonon is created and the photon shifts down in frequency. 
Due to energy and momentum conservation, both terms incorporate a frequency shift ($e^{\pm i \Omega t}$) and momentum shift ($k \pm q$) as depicted in Fig. \ref{fig1}a. 
The momentum shift modifies the original phase matching condition and can enable coupling to a resonator which would otherwise be dark.
Coupling to the resonator, including coupling enabled by the action of synthetic phonons, has a significant impact on wave transmission through the waveguide due to resonant absorption or reflection.
In our proposed system, the steady-state forward transmission coefficient ($S_{21}$) as a function of frequency $\omega$ is evaluated (see the Supplementary Information for a complete derivation) to be
\begin{equation} \label{s21}
\begin{split}
    S_{21} = e^{-i k \ell (N-1)} - \frac{C_2^0 C_1^0 }{i (\omega - \omega_0) + \gamma} - \frac{C_2^- C_1^+}{i (\omega + \Omega - \omega_0) + \gamma} - \frac{C_2^+ C_1^- }{i (\omega - \Omega - \omega_0) + \gamma} ~.
\end{split}
\end{equation}
Here, $S_{21}$ is a linear transfer function and terms corresponding to transmission with a frequency shift have been dropped.
The steady-state backward transmission coefficient ($S_{12}$) is similarly
\begin{equation} \label{s12}
\begin{split}
    S_{12} =  e^{-i k \ell (N-1)} - \frac{C_1^0 C_2^0}{i (\omega - \omega_0) + \gamma} - \frac{C_1^- C_2^+}{i (\omega + \Omega - \omega_0) + \gamma} - \frac{C_1^+ C_2^- }{i (\omega - \Omega - \omega_0) + \gamma} ~.
\end{split}
\end{equation}
From the above equations we find that the synthetic phonon-enabled coupling results in a distinct transmission spectrum where resonant absorption can occur at the original resonance frequency $\omega_0$ as well as the shifted frequencies $\omega_0 \pm \Omega$.
We will hereafter refer to absorption at $\omega_0 - \Omega$ as the Stokes sideband and absorption at $\omega_0 + \Omega$ as the anti-Stokes sideband (see Fig. \ref{fig1}a).
Since the sideband coupling constants are not required to be equal, i.e. $C_1^+ \neq C_2^+$, transmission at these sidebands can be strongly nonreciprocal.
Additionally, we note that absorption at the sidebands is in general asymmetric due to the frequency dependence of $k$.

\vspace{12 pt}

\begin{figure}
\begin{adjustwidth}{-.4in}{-.4in}
    \centering
    \makebox[\textwidth][c]{\includegraphics[width = \linewidth]{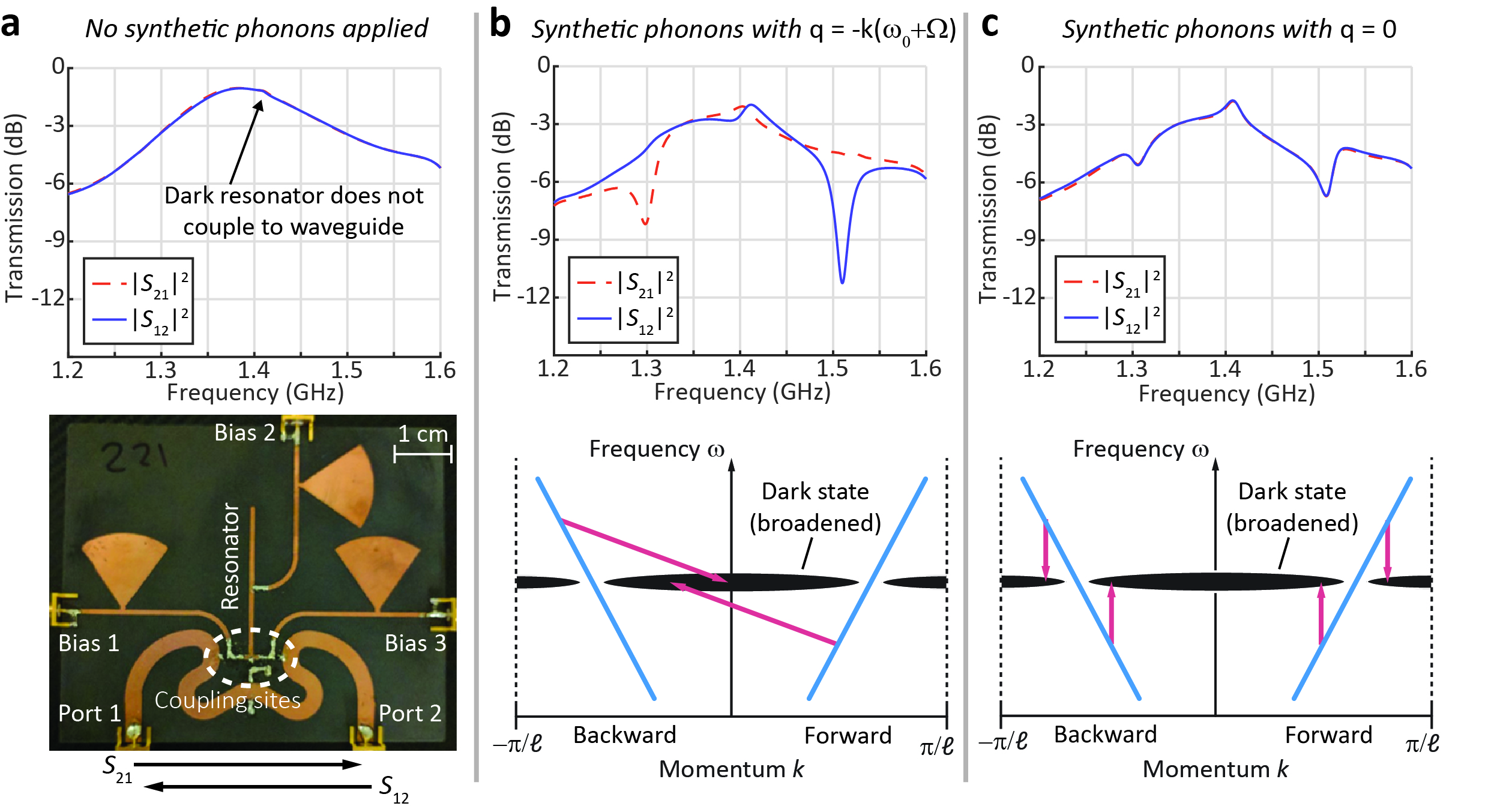}}%
    \caption{Experimentally measured nonreciprocal resonant absorption from synthetic-phonon enabled coupling between a microstrip waveguide and resonator. 
    (a) Picture of the experimental circuit and measured power transmission without synthetic phonon bias applied. The dark resonance does not interact with the waveguide due to destructive interference from multiple coupling sites, and no dip in transmission is observed. 
    (b) Synthetic phonons with momentum $q$ are applied such that $-k(\omega_0 + \Omega) - q = 0$, enabling nonreciprocal coupling to the resonator. Absorption occurs at the Stokes sideband for forward propagation ($S_{21}$ measurement) and the anti-Stokes sideband for reverse propagation ($S_{12}$ measurement). The resonance is broadened in momentum space, since the finite number of coupling sites ($N = 3$) only completely destructively interfere for $k = \pm 2 \pi / 3 \ell$.
    (c) There is no momentum shift when synthetic phonons with zero momentum ($q = 0$) are applied. However, coupling is observed at the sideband frequencies due to waveguide dispersion. The system is reciprocal because these synthetic phonons do not break time-reversal symmetry.
 }
    \label{fig3}
\end{adjustwidth}
\end{figure}

To experimentally validate our theory, we implemented a waveguide-resonator system with three coupling sites ($N = 3$) using a microstrip waveguide and stub resonator (Fig. \ref{fig3}a, bottom). 
The fabricated resonator has a loaded resonant frequency $\omega_0 / 2 \pi \approx 1.4$ GHz. 
The waveguide and resonator are coupled through variable capacitors (varactor diodes) which enable dynamic control over the coupling constants $c_n$ (see Methods).
We design the coupling site separation such that $ k \ell = 2 \pi / 3$ at $\omega = \omega_0$, resulting in a complete phase mismatch and creating a dark state. 
Without a synthetic phonon bias the measured response for this circuit (Fig. \ref{fig3}a, top) does not indicate any dips in transmission corresponding to resonant absorption, confirming that interactions between the resonator and waveguide are suppressed by the phase mismatch. 
The broadband background transmission losses are caused by reflection from and losses in the capacitive coupling network.

We next apply synthetic phonons with frequency $\Omega / 2\pi = 104$ MHz and momentum $q = -k$ at $\omega = \omega_0 + \Omega$, implemented through a phase offset $\theta_n = \frac{5 \pi}{3}  (n-1)$, to the system as described by Eqn. (\ref{modulation}).
This phonon momentum was empirically tuned to maximize the coupling rate $C_1^+ C_2^-$ between the resonator and backward waveguide mode at the anti-Stokes sideband.
From Eq. (\ref{s21}) we see that, neglecting any frequency dependence, $C_1^+ C_2^-$ also describes coupling for the forward waveguide mode at the Stokes sideband.
Thus resonant absorption should occur nonreciprocally: at the anti-Stokes sideband for backward transmission and at the Stokes sideband for forward transmission.
The coupling rate $C_2^+ C_1^-$ is simultaneously minimized by this choice of phonon momentum, so no absorption is expected at these frequencies for the opposite directions (anti-Stokes for forward transmission and Stokes for backward transmission). 
The measured forward ($S_{21}$) and backward ($S_{12}$) transmission coefficients for this system are shown in Fig. \ref{fig3}b.
As predicted, resonant absorption occurs at $\approx 1.3$ GHz only in the forward direction and at $\approx 1.5$ GHz only in the backward direction.
The frequency dependence of $k$ creates a slight phase mismatch at the Stokes sideband, resulting in reduced absorption.
The measured absorption is highly nonreciprocal, no resonant absorption is observed at $\approx 1.3$ GHz in the backward direction or at $\approx 1.5$ GHz in the forward direction, validating that $C_1^- = C_2^+ \approx 0$.
In this experiment, we have shown that synthetic phonons can facilitate coupling to dark states by modifying the original phase matching condition.
Such phonon-assisted coupling results in nonreciprocal transmission if this modified phase matching condition is not satisfied for both directions simultaneously.
However, synthetic phonons that do not modify the phase matching condition, i.e. phonons with zero momentum ($q = 0$), can also enable coupling to a dark resonator due to the frequency dependence of $k$. 
This case, which we show experimentally in Fig. \ref{fig3}c, demonstrates that both waveguide directions can couple to a dark state simultaneously, resulting in reciprocal transmission.
We note that because a partial phase mismatch remains, the coupling is weaker at the sidebands and absorption is reduced compared to Fig. \ref{fig3}b.
%

%%%%%%%%%%%%%%%%%
\subsection*{Demonstration of elementary nonreciprocal devices}
%%%%%%%%%%%%%%%%%

%
Nonreciprocal devices such as isolators and circulators are important tools for controlling wave propagation and have a wide range of uses, from protecting lasers against reflections \cite{What} to facilitating full duplex communications \cite{Reiskarimian}. 
The gyrator, a fundamental nonreciprocal building block that introduces a unidirectional $\pi$ phase shift, can be used to produce a variety of nonreciprocal circuits including isolators and circulators. 
Below, we experimentally show how both isolators and gyrators can be directly created through synthetic phonon-enabled coupling to a single dark state. 
Additionally, in the Supplementary Information we provide preliminary experimental evidence of a four-port circulator implemented using synthetic phonon enabled coupling between a dark state and two waveguides.

\begin{figure}
\begin{adjustwidth}{-.4in}{-.4in}
    \centering
    \makebox[\textwidth][c]{\includegraphics[width = \linewidth]{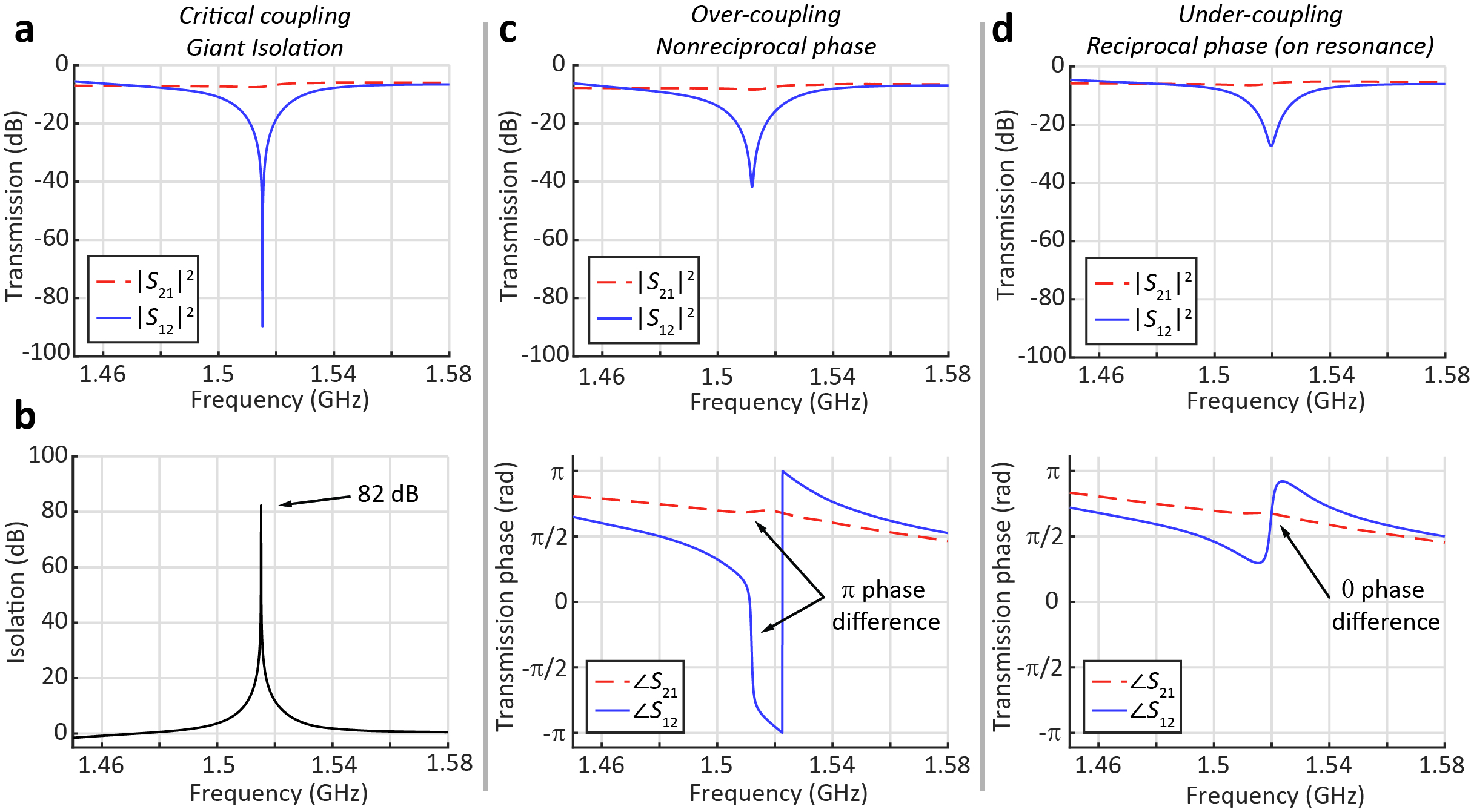}}%
    \caption{Experimental demonstration of essential nonreciprocal functions. Nonreciprocal coupling is enabled by synthetic phonons with the same momentum $q$ but varying amplitude $\delta_c$ in all plots. All measurements are focused on the anti-Stokes sideband.
    (a) Nearly zero transmission ($< -92$ dB) is measured through the waveguide in the direction which is critically coupled ($C_1^+ C_2^- = \gamma$).
    (b) Measured power isolation contrast for the case shown in (a).
    (c) For an over-coupled resonator, there is a $\pi$ nonreciprocal phase shift at the anti-Stokes sideband frequency.
    (d) When the resonator is under-coupled, there is no phase shift at the anti-Stokes sideband frequency. }
    \label{fig4}
\end{adjustwidth}
\end{figure}

We first consider the case of an isolator with high transmission amplitude in the forward direction and zero transmission in the backward direction, operating at the anti-Stokes sideband frequency $\omega_0 + \Omega$.
Examining Eqs. (\ref{s21}) and (\ref{s12}) we find that this case occurs when $C_1^+ C_2^- = \gamma$ (the critical coupling condition \cite{Cai}) and $C_2^+ C_1^- = 0$.
We experimentally investigated this case using the circuit shown in Fig. \ref{fig3}a. 
The resonance frequency was tuned to $\omega_0 / 2\pi \approx 1.42$ GHz, and synthetic phonons were again applied with frequency $\Omega / 2\pi = 104$ MHz and $q = -k(\omega_0 + \Omega)$.
The synthetic phonon amplitude $\delta_c$ was increased until the critical coupling condition $C_1^+ C_2^- = \gamma$ was reached. 
The measured forward ($S_{21}$) and backward ($S_{12}$) transmission coefficients for synthetic phonons with this critical amplitude are presented in Fig. \ref{fig4}a. 
We observe a large Lorentzian dip in the measured backward transmission, which drops to below $-89$ dB at $1.52$ GHz. No resonant absorption is visible in the forward direction.
The measured isolation contrast (Fig. \ref{fig4}b) exceeds $82$ dB with a 10 dB bandwidth of approximately 12 MHz.
We next analyze the case of a gyrator, where high transmission amplitude occurs in both directions, but the backward transmission is phase shifted by $\pi$ in comparison to forward transmission.
Considering the same system as above, it is evident from Eqs. (\ref{s21}) and (\ref{s12}) that this case occurs if the phonon amplitude is increased such that $C_1^+ C_2^- \approx 2 \gamma$ (strong over-coupling) while the opposite direction remains uncoupled.
We experimentally realize nonreciprocal over-coupling by further increasing the synthetic phonon amplitude such that $C_1^+ C_2^- > \gamma$, and observe the anticipated nonreciprocal $\pi$ phase shift at the anti-Stokes sideband frequency $\approx 1.52$ GHz  (Fig. \ref{fig4}b).
Unfortunately, we were unable to realize the required synthetic phonon amplitude to achieve $C_1^+ C_2^- \approx 2 \gamma$ due to limitations caused by non-linearity in the varactor diodes. 
For comparison, we also show measured transmission amplitude and phase for the under-coupled case, where $C_1^+ C_2^- < \gamma$ and there is no nonreciprocal $\pi$ phase shift at the anti-Stokes sideband.
\subsection*{Higher-order nonreciprocal transfer functions}
\begin{figure}
\begin{adjustwidth}{-.4in}{-.4in}
    \centering
    \makebox[\textwidth][c]{\includegraphics[width = \linewidth]{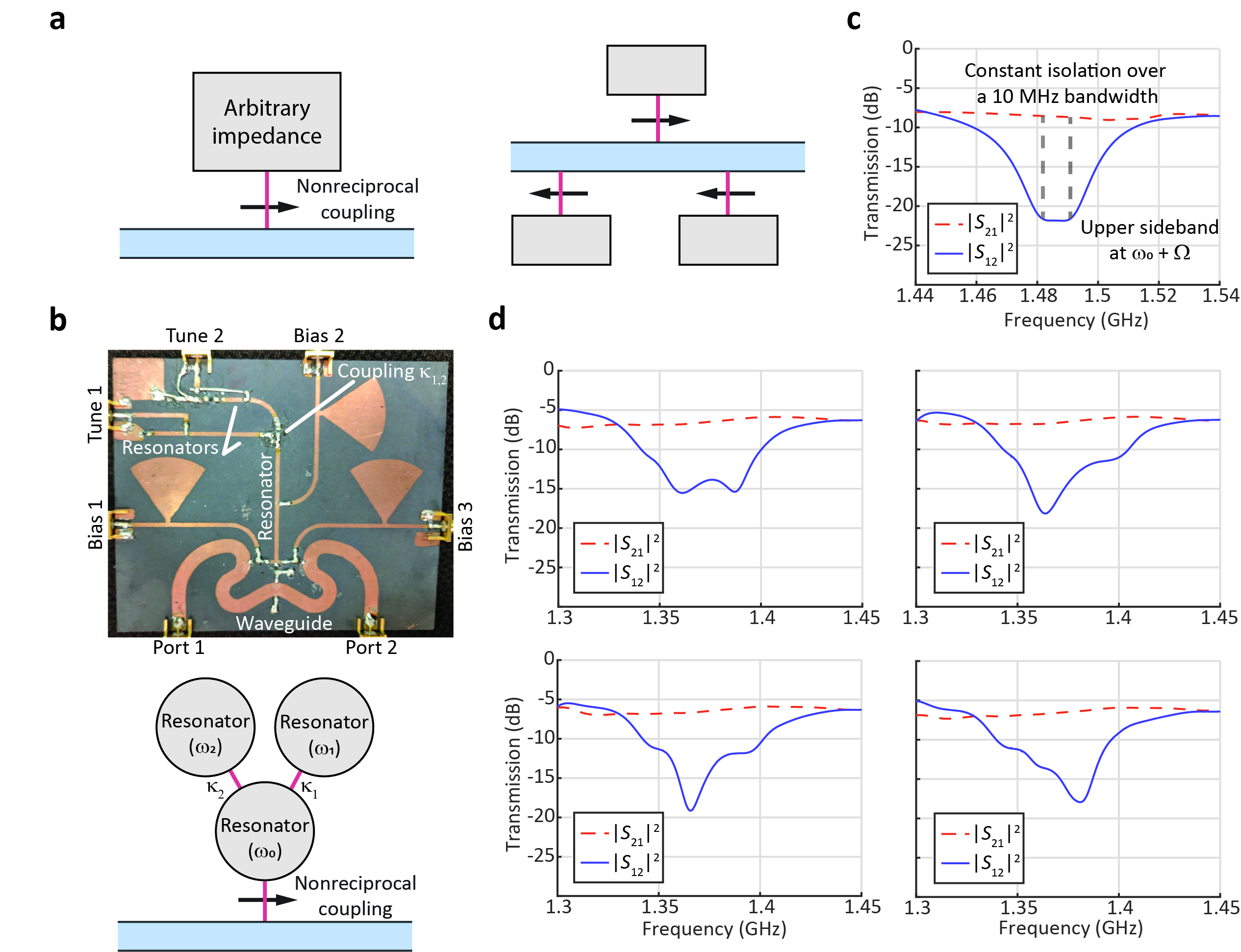}}%
    \caption{Higher-order nonreciprocal transfer functions created through non-reciprocal coupling to resonator networks.
    (a) Nonreciprocal coupling can be engaged to an arbitrary band-limited impedance network. Several of these impedance networks can be simultaneously coupled in either direction to create customizable responses.
    (b) Photograph of the circuit use to demonstrate customizable nonreciprocal transfer functions.
    (b) Schematic of the circuit from (a).
    (c) Measured power transmission showing a flat band over which a constant isolation response is obtained.
    (d) Experimental demonstration of four distinct nonreciprocal transfer functions obtained by tuning the resonator network.
    }
    \label{fig5}
\end{adjustwidth}
\end{figure}

While high-order filters are often necessary for signal processing applications \cite{Little, Hryniewicz}, a platform for integrating such functionality into non-magnetic nonreciprocal systems has not yet been shown.
Frequency-selective nonreciprocal devices in literature have been mainly limited to Lorentizian shaped transfer functions \cite{Lira, Fleury, Estep, Kim, Dong, EstepIEEE, Reiskarimian}. 
%
%Higher-order nonreciprocal filters have potential for use in any signal processing application which requires a specific frequency response that differs depending on the direction of propagation.
%
Synthetic phonon-enabled coupling is a technique uniquely suited to address this challenge because it can permit unidirectional access to arbitrary band-limited load impedances (Fig. \ref{fig5}a), producing arbitrary nonreciprocal responses.
Additionally, different frequency responses could be simultaneously achieved in opposite directions by coupling each direction to an appropriate resonator network \cite{Orta}.
Implementing this idea, we experimentally demonstrate non-Lorentzian nonreciprocal transfer functions using the circuit shown in Fig. \ref{fig5}b, which is a modified version of that in Fig. \ref{fig3}a. 
Here, two additional microstrip stub resonators with tunable resonance frequencies are coupled to the original stub resonator used in previous experiments (Fig. \ref{fig5}b), providing six additional degrees of freedom: the additional resonance frequencies $\omega_1,~\omega_2$, inter-resonator coupling rates $\kappa_1,~\kappa_2$, and linewidths $\gamma_1,~\gamma_2$.
A maximally flat nonreciprocal filter with constant isolation over an appreciable bandwidth is arguably one of the most important functionalities that cannot be implemented using a single resonant response.
Such a flat response can be approximated in the three resonator network using $\gamma_1 = \gamma_2 = \gamma$, $\kappa_1 = \kappa_2 = \frac{9}{14} \gamma$, $\omega_1 = \omega_0 + \frac{3}{7}\gamma$, and $\omega_2 = \omega_0 - \frac{3}{7} \gamma$, where the loss rate $\gamma$ and resonance frequency $\omega_0$ are associated with the original resonator.
We empirically tuned both the resonance frequencies ($\omega_1$, $\omega_2$) and coupling rates ($\kappa_1, \kappa_2$) of the additional resonators in our circuit (Fig. \ref{fig5}b) near these values until the desired transfer function was achieved.
Since the resonators are fabricated on the same substrate and conductor, their linewidths are intrinsically equal.
The experimentally measured transmission through the waveguide (Fig. \ref{fig5}c) exhibits nearly constant isolation of 14 dB over a 10 MHz bandwidth. 
In Figure \ref{fig5}d we present four additional examples of arbitrary nonreciprocal transfer functions obtained by varying the inter-resonator coupling strength and frequency separation of the three resonators.
In these experiments we observe consistently flat forward transmission ($S_{21}$) even though the reverse transmission ($S_{12}$) varies, clearly demonstrating that propagation in the uncoupled direction is largely unaffected by changes to the impedance network.
%
%
%
%
%
%
%%%%%%%%%%%%%%%%%%%%%%%%%%%%%%%
\section*{Discussion}
%%%%%%%%%%%%%%%%%%%%%%%%%%%%%%%

%
In this work we have demonstrated that coupling to arbitrary networks of resonators can be engaged nonreciprocally, and have used such coupling to realize new higher-order nonreciprocal filters as well as fundamental nonreciprocal devices.
Furthermore, we have introduced synthetic phonons with a precisely controlled momentum, which can replicate the action of optically active phonons without relying on any dispersion relation.
%
%There are many potential improvements or modifications that can be made to the devices presented here, namely increasing the Q-factor of the resonator so as to improve the ratio $C_1^+ C_2^- / \gamma$ and achieve near lossless gyration.
%
%Higher Q-factor resonators would also be useful in the creation of devices with other capabilities, such as high-order nonreciprocal filters with sharp features and circulators with high contrast.
%
Although our experiments take place in microstrip circuits, the nonreciprocal behavior of our system is captured by coupled-mode theory and thus the underlying method can be extended to a wide variety of physical systems. 
Additionally, the concept of nonreciprocal coupling can be applied broadly to a number of band-limited devices besides resonators, including antennas, amplifiers, oscillators, and sensors, allowing the creation of highly customizable integrated devices.
%

%
%
%
%
%
%
%%%%%%%%%%%%%%%%%%%%%%%%%%%%%%%
%%%%%%%%%%%%%%%%%%%%%%%%%%%%%%%
\section*{Methods}
%%%%%%%%%%%%%%%%%%%%%%%%%%%%%%%
%%%%%%%%%%%%%%%%%%%%%%%%%%%%%%%

%%%%%%%%%%%%%%%%%%%%%%%%%%%%%%%
\subsection*{Experimental Setup}
%%%%%%%%%%%%%%%%%%%%%%%%%%%%%%%

Our microwave circuits are fabricated on Rogers RT/duroid 5880 substrate with a 1 oz copper conductor, and consist of a microstrip waveguide coupled to a ring resonator by $N$ varactor diodes (Skyworks SMV1275) that act as variable capacitors. 
The coupling strength $c_n$ of each capacitive coupler is an approximately linear function of the applied voltage $V_n$ (for small changes), allowing a modulation of $c_n$ that is proportional to a modulation of $V_n$. 
We first apply a DC bias to each varactor diode with a DC power supply (Agilent E3631A), which lowers the capacitance and decreases the background reflection caused by the coupling system.
On top of this bias, we apply a small sinusoidal signal from a signal generator (HP-8647B), which is set at a frequency of 104 MHz. This signal is split (Minicircuits ZA3CS-400-3W-S) into three variable phase shifters (Minicircuits JSPHS-150+) so that the phase shift between each signal can be independently controlled. 
The DC bias and three 104 MHz modulation signals are combined through a three bias tees (Minicircuits ZFBT-4R2GW-FT+) and connected directly to the circuit, through the ports labeled \textit{Bias} in Fig. \ref{fig3}.
On each circuit, butterfly band-pass filters were incorporated along with a another low-frequency biasing tee (Johanson Technology L-14C10N-V4T 10 nH inductor and Johanson Technology R14S 6.8 pF capacitor) as shown in Fig. \ref{fig3} to isolate the lower frequency (104 MHz) bias modulation from the higher frequency ($\approx 1.4$ GHz) resonant circuit.

%%%%%%%%%%%%%%%%%%%%%%%%%%%%%%%
\subsection*{Measurement}
%%%%%%%%%%%%%%%%%%%%%%%%%%%%%%%

We measured the transmission parameters ($S_{21}$, $S_{12}$) of the circuit using a Keysight E5063A vector network analyzer. The network analyzer measurement was calibrated to the ends of the SMA cables that connected to the surface-mount SMA connectors on the circuit board. Thus, the data presented throughout this paper only reflect the S parameters of the circuits that we have developed, while eliminating any parasitic effects from the cables and supporting systems.

\section*{Acknowledgments}
We acknowledge funding support from the US Office of Naval Research Director of Research Early Career Grant (grant N00014-16-1-2830), the National Science Foundation Emerging Frontiers in Research and Innovation NewLAW program (grant EFMA-1627184), and a National Science Foundation Graduate Research Fellowship.

\section*{Author Contributions}
C.W.P. and G.B. conceived the idea and with S.K. developed the theory.  C.W.P. designed and carried out the experiments.  J.T.B. provided material support and guidance for the experiments.  All authors analyzed the data and co-wrote the paper.  G.B. supervised all aspects of this project.

\vspace{24pt}

\small{\bibliography{references.bib}}

\begin{thebibliography}{10}
\expandafter\ifx\csname url\endcsname\relax
  \def\url#1{\texttt{#1}}\fi
\expandafter\ifx\csname urlprefix\endcsname\relax\def\urlprefix{URL }\fi
\providecommand{\bibinfo}[2]{#2}
\providecommand{\eprint}[2][]{\url{#2}}

\bibitem{Boyd}
\bibinfo{author}{Boyd, R.~W.}
\newblock \emph{\bibinfo{title}{Nonlinear optics}}
  (\bibinfo{publisher}{Academic press}, \bibinfo{year}{2003}).

\bibitem{Bahl2}
\bibinfo{author}{Bahl, G.}, \bibinfo{author}{Tomes, M.},
  \bibinfo{author}{Marquardt, F.} \& \bibinfo{author}{Carmon, T.}
\newblock \bibinfo{title}{Observation of spontaneous {Brillouin} cooling}.
\newblock \emph{\bibinfo{journal}{Nature Physics}}
  \textbf{\bibinfo{volume}{8}}, \bibinfo{pages}{203--207}
  (\bibinfo{year}{2012}).

\bibitem{SKim}
\bibinfo{author}{Kim, S.} \& \bibinfo{author}{Bahl, G.}
\newblock \bibinfo{title}{Role of optical density of states in {Brillouin}
  optomechanical cooling}.
\newblock \emph{\bibinfo{journal}{Optics Express}}
  \textbf{\bibinfo{volume}{25}}, \bibinfo{pages}{776--784}
  (\bibinfo{year}{2017}).

\bibitem{SKim2}
\bibinfo{author}{Kim, S.}, \bibinfo{author}{Xu, X.}, \bibinfo{author}{Taylor,
  J.~M.} \& \bibinfo{author}{Bahl, G.}
\newblock \bibinfo{title}{Dynamically induced robust phonon transport and
  chiral cooling in an optomechanical system}.
\newblock \emph{\bibinfo{journal}{Nature Communications}}
  \textbf{\bibinfo{volume}{8}}, \bibinfo{pages}{205} (\bibinfo{year}{2017}).

\bibitem{Hwang}
\bibinfo{author}{Hwang, I.~K.}, \bibinfo{author}{Yun, S.~H.} \&
  \bibinfo{author}{Kim, B.~Y.}
\newblock \bibinfo{title}{All-fiber-optic nonreciprocal modulator}.
\newblock \emph{\bibinfo{journal}{Optics Letters}}
  \textbf{\bibinfo{volume}{22}}, \bibinfo{pages}{507--509}
  (\bibinfo{year}{1997}).

\bibitem{Yu}
\bibinfo{author}{Yu, Z.} \& \bibinfo{author}{Fan, S.}
\newblock \bibinfo{title}{Complete optical isolation created by indirect
  interband photonic transisions}.
\newblock \emph{\bibinfo{journal}{Nature Photonics}}
  \textbf{\bibinfo{volume}{3}}, \bibinfo{pages}{91--94} (\bibinfo{year}{2008}).

\bibitem{Bahl}
\bibinfo{author}{Bahl, G.}, \bibinfo{author}{Zehnpfennig, J.},
  \bibinfo{author}{Tomes, M.} \& \bibinfo{author}{Carmon, T.}
\newblock \bibinfo{title}{Stimulated optomechanical excitation of surface
  acoustic waves in a microdevice}.
\newblock \emph{\bibinfo{journal}{Nature Communications}}
  \textbf{\bibinfo{volume}{2}} (\bibinfo{year}{2011}).

\bibitem{Pant}
\bibinfo{author}{Pant, R.} \emph{et~al.}
\newblock \bibinfo{title}{On-chip stimulated {Brillouin} scattering}.
\newblock \emph{\bibinfo{journal}{Optics Express}}
  \textbf{\bibinfo{volume}{19}}, \bibinfo{pages}{8285--8290}
  (\bibinfo{year}{2011}).

\bibitem{Kittlaus}
\bibinfo{author}{Kittlaus, E.~A.}, \bibinfo{author}{Otterstrom, N.~T.} \&
  \bibinfo{author}{Rakich, P.~T.}
\newblock \bibinfo{title}{On-chip inter-modal brillouin scattering}.
\newblock \emph{\bibinfo{journal}{Nature Communications}}
  \textbf{\bibinfo{volume}{8}}, \bibinfo{pages}{15819} (\bibinfo{year}{2017}).

\bibitem{Kamal}
\bibinfo{author}{Kamal, A.}, \bibinfo{author}{Clarke, J.} \&
  \bibinfo{author}{Devoret, M.~H.}
\newblock \bibinfo{title}{Noiseless non-reciprocity in a parametric active
  device}.
\newblock \emph{\bibinfo{journal}{Nature Physics}}
  \textbf{\bibinfo{volume}{7}}, \bibinfo{pages}{311--315}
  (\bibinfo{year}{2011}).

\bibitem{Kang}
\bibinfo{author}{Kang, M.~S.}, \bibinfo{author}{Butsch, A.} \&
  \bibinfo{author}{Russell, P. S.~J.}
\newblock \bibinfo{title}{Reconfigurable light-driven opto-acoustic isolators
  in photonic crystal fibre}.
\newblock \emph{\bibinfo{journal}{Nature Photonics}}
  \textbf{\bibinfo{volume}{5}}, \bibinfo{pages}{549--553}
  (\bibinfo{year}{2011}).

\bibitem{Hafezi}
\bibinfo{author}{Hafezi, M.} \& \bibinfo{author}{Rabl, P.}
\newblock \bibinfo{title}{Optomechanically induced non-reciprocity in microring
  resonators}.
\newblock \emph{\bibinfo{journal}{Optics Express}}
  \textbf{\bibinfo{volume}{20}}, \bibinfo{pages}{7672--7684}
  (\bibinfo{year}{2012}).

\bibitem{Poulton}
\bibinfo{author}{Poulton, C.~G.} \emph{et~al.}
\newblock \bibinfo{title}{Design for broadband on-chip isolator using
  stimulated brillouin scattering in dispersion-engineered chalcogendide
  waveguides}.
\newblock \emph{\bibinfo{journal}{Optics Express}}
  \textbf{\bibinfo{volume}{20}}, \bibinfo{pages}{21235--21246}
  (\bibinfo{year}{2012}).

\bibitem{Lira}
\bibinfo{author}{Lira, H.}, \bibinfo{author}{Yu, Z.}, \bibinfo{author}{Fan, S.}
  \& \bibinfo{author}{Lipson, M.}
\newblock \bibinfo{title}{Electrically driven nonreciprocity induced by
  interband photonic transition on a silicon chip}.
\newblock \emph{\bibinfo{journal}{Physical Review Letters}}
  \textbf{\bibinfo{volume}{109}}, \bibinfo{pages}{033901}
  (\bibinfo{year}{2012}).

\bibitem{Estep}
\bibinfo{author}{Estep, N.~A.}, \bibinfo{author}{Sounas, D.~L.},
  \bibinfo{author}{Soric, J.} \& \bibinfo{author}{Alu, A.}
\newblock \bibinfo{title}{Magnetic-free non-reciprocity and isolation based on
  parametrically modulated coupled-resonator loops}.
\newblock \emph{\bibinfo{journal}{Nature Physics}}
  \textbf{\bibinfo{volume}{10}}, \bibinfo{pages}{923--927}
  (\bibinfo{year}{2014}).

\bibitem{Qin}
\bibinfo{author}{Qin, S.}, \bibinfo{author}{Xu, Q.} \& \bibinfo{author}{Wang,
  Y.~E.}
\newblock \bibinfo{title}{Nonreciprocal components based on distributed
  modulated capacitors}.
\newblock \emph{\bibinfo{journal}{IEEE Transactions on Microwave Theory and
  Techniques}} \textbf{\bibinfo{volume}{62}}, \bibinfo{pages}{2260--2272}
  (\bibinfo{year}{2014}).

\bibitem{Kim}
\bibinfo{author}{Kim, J.}, \bibinfo{author}{Kuzyk, M.~C.},
  \bibinfo{author}{Han, K.}, \bibinfo{author}{Wang, H.} \&
  \bibinfo{author}{Bahl, G.}
\newblock \bibinfo{title}{Non-reciprocal {Brillouin} scattering induced
  transparency}.
\newblock \emph{\bibinfo{journal}{Nature Physics}}
  \textbf{\bibinfo{volume}{11}}, \bibinfo{pages}{275--280}
  (\bibinfo{year}{2015}).

\bibitem{Dong}
\bibinfo{author}{Dong, C.-H.} \emph{et~al.}
\newblock \bibinfo{title}{Brillouin-scattering-induced transparency and
  non-reciprocal light storage}.
\newblock \emph{\bibinfo{journal}{Nature Communications}}
  \textbf{\bibinfo{volume}{6}}, \bibinfo{pages}{6193} (\bibinfo{year}{2015}).

\bibitem{EstepIEEE}
\bibinfo{author}{Estep, N.~A.}, \bibinfo{author}{Sounas, D.~L.} \&
  \bibinfo{author}{Alu, A.}
\newblock \bibinfo{title}{Magnetless microwave circulators based on
  spatiotemporally modulated rings of coupled resonators}.
\newblock \emph{\bibinfo{journal}{IEEE Transactions on Microwave Theory and
  Techniques}} \textbf{\bibinfo{volume}{64}}, \bibinfo{pages}{502--518}
  (\bibinfo{year}{2016}).

\bibitem{Shen}
\bibinfo{author}{Shen, Z.} \emph{et~al.}
\newblock \bibinfo{title}{Experimental realization of optomechanically induced
  non-reciprocity}.
\newblock \emph{\bibinfo{journal}{Nature Photonics}}
  \textbf{\bibinfo{volume}{10}}, \bibinfo{pages}{657--661}
  (\bibinfo{year}{2016}).

\bibitem{Kim2}
\bibinfo{author}{Kim, J.}, \bibinfo{author}{Kim, S.} \& \bibinfo{author}{Bahl,
  G.}
\newblock \bibinfo{title}{Complete linear optical isolation at the microscale
  with ultralow loss}.
\newblock \emph{\bibinfo{journal}{Sci. Rep.}} \textbf{\bibinfo{volume}{7}},
  \bibinfo{pages}{1647} (\bibinfo{year}{2017}).

\bibitem{Shii}
\bibinfo{author}{Shi, Y.}, \bibinfo{author}{Han, S.} \& \bibinfo{author}{Fan,
  S.}
\newblock \bibinfo{title}{Optical circulation and isolation based on indirect
  photonic transitions of guided resonance modes}.
\newblock \emph{\bibinfo{journal}{ACS Photonics}} \textbf{\bibinfo{volume}{4}},
  \bibinfo{pages}{1639--1645} (\bibinfo{year}{2017}).

\bibitem{Fleury}
\bibinfo{author}{Fleury, R.}, \bibinfo{author}{Sounas, D.~L.},
  \bibinfo{author}{Sieck, C.~F.}, \bibinfo{author}{Haberman, M.~R.} \&
  \bibinfo{author}{Alu, A.}
\newblock \bibinfo{title}{Sound isolation and giant linear nonreciprocity in a
  compact acoustic circulator}.
\newblock \emph{\bibinfo{journal}{Science}} \textbf{\bibinfo{volume}{343}},
  \bibinfo{pages}{516--519} (\bibinfo{year}{2014}).

\bibitem{Casimir}
\bibinfo{author}{Casimir, H. B.~G.}
\newblock \bibinfo{title}{Reciprocity theorems and irreversible processes}.
\newblock \emph{\bibinfo{journal}{Proceedings of the IEEE}}
  \textbf{\bibinfo{volume}{51}}, \bibinfo{pages}{1570--1573}
  (\bibinfo{year}{1963}).

\bibitem{Boller}
\bibinfo{author}{Boller, K.-J.}, \bibinfo{author}{Imamoglu, A.} \&
  \bibinfo{author}{Harris, S.~E.}
\newblock \bibinfo{title}{Observation of electromagnetically induced
  transparency}.
\newblock \emph{\bibinfo{journal}{Phys. Rev. Lett.}}
  \textbf{\bibinfo{volume}{66}}, \bibinfo{pages}{2593--2596}
  (\bibinfo{year}{1991}).

\bibitem{Arimondo}
\bibinfo{author}{Arimondo, E.}
\newblock \bibinfo{title}{Coherent population trapping in laser spectroscopy}.
\newblock In \emph{\bibinfo{booktitle}{Progress in Optics}},
  vol.~\bibinfo{volume}{35}, \bibinfo{pages}{257--354}
  (\bibinfo{publisher}{Elsevier, Amsterdam}, \bibinfo{year}{1996}).
\newblock \bibinfo{note}{JILA Pub. 5423}.

\bibitem{OptomechDark}
\bibinfo{author}{Dong, C.}, \bibinfo{author}{Fiore, V.},
  \bibinfo{author}{Kuzyk, M.~C.} \& \bibinfo{author}{Wang, H.}
\newblock \bibinfo{title}{Optomechanical dark mode}.
\newblock \emph{\bibinfo{journal}{Science}} \textbf{\bibinfo{volume}{338}},
  \bibinfo{pages}{1609--1613} (\bibinfo{year}{2012}).

\bibitem{Guo}
\bibinfo{author}{Guo, Q.-H.}, \bibinfo{author}{Kang, M.}, \bibinfo{author}{Li,
  T.-F.}, \bibinfo{author}{Cui, H.-X.} \& \bibinfo{author}{Chen, J.}
\newblock \bibinfo{title}{Slow light from sharp dispersion by exciting dark
  photonic angular momentum states}.
\newblock \emph{\bibinfo{journal}{Optics Letters}}
  \textbf{\bibinfo{volume}{38}}, \bibinfo{pages}{250--252}
  (\bibinfo{year}{2013}).

\bibitem{Yang2}
\bibinfo{author}{Yang, M.} \emph{et~al.}
\newblock \bibinfo{title}{Manipulation of dark photonic angular momentum states
  via magneto-optical effect for tunable slow-light performance}.
\newblock \emph{\bibinfo{journal}{Optics Express}}
  \textbf{\bibinfo{volume}{21}}, \bibinfo{pages}{25035--25044}
  (\bibinfo{year}{2013}).

\bibitem{Haus}
\bibinfo{author}{Haus, H.}
\newblock \emph{\bibinfo{title}{Waves and Fields in Optoelectronics}}
  (\bibinfo{publisher}{Prentice-Hall}, \bibinfo{address}{Englewood Cliffs, New
  Jersey}, \bibinfo{year}{1984}).

\bibitem{CMT}
\bibinfo{author}{Suh, W.}, \bibinfo{author}{Wang, Z.} \& \bibinfo{author}{Fan,
  S.}
\newblock \bibinfo{title}{Temporal coupled-mode theory and the presence of
  non-orthogonal modes in lossless multimode cavities}.
\newblock \emph{\bibinfo{journal}{IEEE Journal of Quantum Electronics}}
  \textbf{\bibinfo{volume}{40}}, \bibinfo{pages}{1511--1518}
  (\bibinfo{year}{2004}).

\bibitem{What}
\bibinfo{author}{Jalas, D.} \emph{et~al.}
\newblock \bibinfo{title}{What is - and what is not - and optical isolator}.
\newblock \emph{\bibinfo{journal}{Nature Photonics}}
  \textbf{\bibinfo{volume}{7}}, \bibinfo{pages}{579--582}
  (\bibinfo{year}{2013}).

\bibitem{Reiskarimian}
\bibinfo{author}{Reiskarimian, N.} \& \bibinfo{author}{Krishnaswamy, H.}
\newblock \bibinfo{title}{Magnetic-free non-reciprocity based on staggered
  commutation}.
\newblock \emph{\bibinfo{journal}{Nature Communications}}
  \textbf{\bibinfo{volume}{7}}, \bibinfo{pages}{11217} (\bibinfo{year}{2016}).

\bibitem{Cai}
\bibinfo{author}{Cai, M.}, \bibinfo{author}{Painter, O.} \&
  \bibinfo{author}{Vahala, K.~J.}
\newblock \bibinfo{title}{Observation of critical coupling in a fiber taper to
  a silica-microsphere whispering-gallery mode system}.
\newblock \emph{\bibinfo{journal}{Phys. Rev. Lett.}}
  \textbf{\bibinfo{volume}{85}}, \bibinfo{pages}{74--77}
  (\bibinfo{year}{2000}).

\bibitem{Little}
\bibinfo{author}{Little, B.~E.}, \bibinfo{author}{Chu, S.~T.},
  \bibinfo{author}{Haus, H.~A.}, \bibinfo{author}{Foresi, J.} \&
  \bibinfo{author}{Laine, J.-P.}
\newblock \bibinfo{title}{Microring resonator channel dropping filters}.
\newblock \emph{\bibinfo{journal}{Journal of Lightwave Technology}}
  \textbf{\bibinfo{volume}{15}}, \bibinfo{pages}{998--1005}
  (\bibinfo{year}{1997}).

\bibitem{Hryniewicz}
\bibinfo{author}{Hryniewicz, J.~V.}, \bibinfo{author}{Absil, P.~P.},
  \bibinfo{author}{Little, B.~E.}, \bibinfo{author}{Wilson, R.~A.} \&
  \bibinfo{author}{Ho, P.-T.}
\newblock \bibinfo{title}{Higher order filter response in coupled microring
  resonators}.
\newblock \emph{\bibinfo{journal}{IEEE Photonics Tech. Lett.}}
  \textbf{\bibinfo{volume}{12}}, \bibinfo{pages}{320--322}
  (\bibinfo{year}{2000}).

\bibitem{Orta}
\bibinfo{author}{Orta, R.}, \bibinfo{author}{Savi, P.},
  \bibinfo{author}{Tascone, R.} \& \bibinfo{author}{Trinchero, D.}
\newblock \bibinfo{title}{Synthesis of multiple-ring-resonator filters for
  optical systems}.
\newblock \emph{\bibinfo{journal}{IEEE Photonics Tech. Lett.}}
  \textbf{\bibinfo{volume}{7}}, \bibinfo{pages}{1447--1449}
  (\bibinfo{year}{1995}).

\end{thebibliography}

\newpage

\newcommand{\beginsupplement}{%
        \setcounter{table}{0}
        \renewcommand{\thetable}{S\arabic{table}}%
        \setcounter{figure}{0}
        \renewcommand{\thefigure}{S\arabic{figure}}%
        \setcounter{equation}{0}
        \renewcommand{\theequation}{S\arabic{equation}}%\
        \setcounter{section}{0}
        \renewcommand{\thesection}{S\arabic{section}}%
}

\beginsupplement

\begin{center}
\Large{\textbf{Supplementary Information: \\ Synthetic phonons enable nonreciprocal coupling to dark states}} \\
\vspace{12pt}
\vspace{12pt}
\large{{Christopher W. Peterson}$^1$,
{Seunghwi Kim}$^2$,
{Jennifer T Bernhard}$^1$,
\mbox{Gaurav Bahl}$^{2,\ast}$} \\
\vspace{12pt}
{\footnotesize{University of Illinois at Urbana-Champaign\\}}
{\footnotesize{$^1$Department of Electrical and Computer Engineering\\}}
{\footnotesize{$^2$Department of Mechanical Science and Engineering\\}}
{\footnotesize{$^*$To whom correspondence should be addressed; bahl@illinois.edu}} \\
%\vspace{12pt}
\vspace{12pt}
\end{center}

%\maketitle
%
\section{Steady-state transmission coefficients}
\begin{figure}
    \centering
    \includegraphics [width=0.75\linewidth] {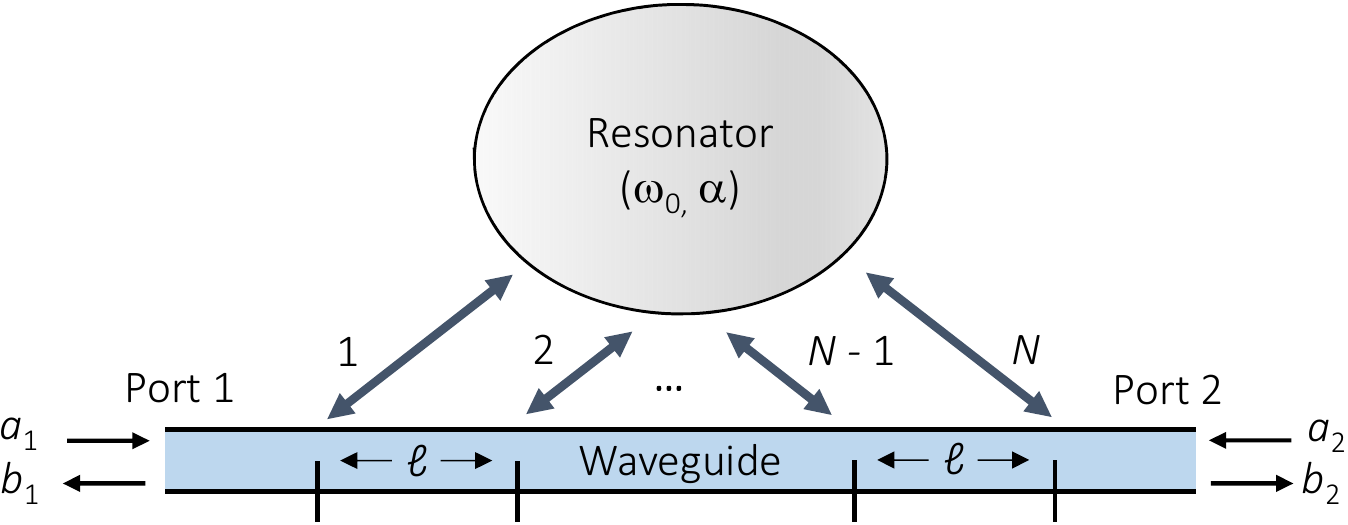}
    \caption{
    Schematic of a waveguide-resonator system with one resonance and two ports. The input and output at port $m$ is described by $a_m$ and $b_m$ respectively and the resonant field is described by $\alpha$. The ports are coupled directly via a waveguide and are also coupled to a resonant modes using a system of $N$ spatiotemporally modulated coupling sites with coupling rate $c_n$. The individual sites are spatially distributed over the waveguide with subwavelength spacing $\ell$ in order to generate sensitivity to wave propagation direction.
    }
    \label{figS1}
\end{figure}

In this section we explicitly derive the nonreciprocal scattering parameters for a waveguide resonator system with multiple spatiotemporally modulated couplers. The system under consideration, shown in Fig. \ref{figS1}, consists of two ports connected by a waveguide with input and output fields described by $a_m$ and $b_m$ (where $m$ is the port number) and a resonant mode with angular frequency $\omega_0$ and amplitude $\alpha$. 
The mathematical derivations in this paper are an extension of temporal coupled-mode theory \cite{CMT} modified to incorporate time-varying couplers. We begin with the differential equation describing the resonant field $\alpha$ of a standing-wave resonator in the time-domain with inputs $a_1$ and $a_2$ that are time-harmonic with frequency $\omega$. 
\begin{equation} \label{S1}
    \dot{\alpha} = (i \omega_0 - \gamma) \alpha + i C_1 a_1 + i C_2 a_2 ~ .
\end{equation}
This equation can be modified for a traveling wave resonator (such as an optical whispering-gallery mode resonator) by splitting $\alpha$ into two orthogonal modes which are each only coupled to a single port.
\begin{equation} \label{ST}
    \begin{split}
        & \dot{\alpha_1} = (i \omega_0 - \gamma) \alpha_1 + i C_1 a_1 ~ , \\
        & \dot{\alpha_2} = (i \omega_0 - \gamma) \alpha_2 + i C_2 a_2 ~ .
    \end{split}
\end{equation}
We can express the output fields $b_2$, $b_1$ as
\begin{equation} \label{S2}
    \begin{split}
        & b_2 = e^{-i \beta \ell (N-1)} a_1 + i C_2 \alpha_{(1)} ~ , \\
        & b_1 = e^{-i \beta \ell (N-1)} a_2 + i C_1 \alpha_{(2)} ~ ,
   \end{split}
\end{equation}
where the subscript of $\alpha$ is used only for the traveling wave resonator case. In these equations $\omega_0$ is the resonance frequency, $\gamma$ is the decay rate of the mode, and $C_m$ is the effective coupling between port $m$ and the resonant mode. We assume the waveguide is matched to the ports and is lossless with a propagation constant $k$. The couplers are spatially separated in the waveguide by a length $\ell$, which leads to the term $e^{-i k \ell (N-1)}$ to account for the phase associated with a wave traveling between ports. At steady state the solution to Eqn. (\ref{S1}) (a standing wave resonator) is 
\begin{equation} \label{S1S}
    \alpha = i \frac{C_1 a_1 + C_2 a_2}{i (\omega - \omega_0) + \gamma} ~ .
\end{equation}
We can then rewrite Eqn. (\ref{S2}) as
\begin{equation} \label{S3}
    \begin{split}
        & b_2 = \left( e^{-i \beta \ell (N-1)} a_1 - \frac{C_2 C_1 a_1 + C_2 C_2 a_2}{i (\omega - \omega_0) + \gamma} \right) ~ , \\
        & b_1 = \left( e^{-i \beta \ell (N-1)} a_2 - \frac{C_1 C_1 a_1 + C_1 C_2 a_2}{i (\omega - \omega_0) + \gamma} \right) ~ .
    \end{split}
\end{equation}
In the system under consideration, shown in Fig. \ref{figS1}, the effective coupling rates $C_1$ and $C_2$ can be written as
\begin{equation} \label{S4}
    \begin{split}
        & C_1 = \sum_{n=1}^N c_n e^{-i k \ell (n-1)} ~ , \\
        & C_2 = \sum_{n=1}^N c_n e^{i k \ell (n-1)} ~ ,
    \end{split}
\end{equation}
%    c_n = c_0 + c_M \cos(\Omega t - q \ell (n-1)) ~
where we modulate the coupling rate of each individual coupler $c_n$ with the function $_n = c_0 + \delta_c \cos \big( \Omega t - q \ell (n-1) \big) $. Under this modulation the effective coupling rates $C_1$ and $C_2$ can be expressed as
\begin{equation} \label{Smod_coupling1}
\begin{split}
   C_1 = \overbrace{ c_0 \sum_{n=1}^{N} e^{-i k \ell (n-1)} }^{C_1^0}  + \overbrace{ \frac{\delta_c}{2} e^{i \Omega t} \sum_{n=1}^{N} e^{- i (k + q) \ell (n-1)} }^{C_1^+} 
    + \overbrace{ \frac{\delta_c}{2} e^{-i \Omega t} \sum_{n=1}^{N} e^{-i (k - q) \ell (n-1)} }^{C_1^-} ~,
\end{split}
\end{equation}
\begin{equation} \label{Smod_coupling2}
\begin{split}
   C_2 = \overbrace{ c_0 \sum_{n=1}^{N} e^{i k \ell (n-1)} }^{C_2^0} + \overbrace{ \frac{\delta_c}{2} e^{i \Omega t} \sum_{n=1}^{N} e^{i (k - q) \ell (n-1)} }^{C_2^+} 
   + \overbrace{ \frac{\delta_c}{2} e^{-i \Omega t} \sum_{n=1}^{N} e^{i (k + q) \ell (n-1)} }^{C_2^-} ~.
\end{split}
\end{equation}
\vspace{11pt} \par
For simplicity we assume a resonant mode with only one input ($a_1 \neq 0 ,~ a_2 = 0)$, which could be a standing wave or traveling wave resonator. The steady-state field in the resonator is thus 
\begin{equation} \label{S6}
    \begin{split}
        \alpha & = a_1 \Big[ \frac{ i C_1^0 }{i (\omega - \omega_0) + \gamma} + \frac{ i C_1^+ }{i (\omega + \Omega - \omega_0) + \gamma} + \frac{ i C_1^- }{i (\omega - \Omega - \omega_0) + \gamma} \Big] ~ .
    \end{split}
\end{equation}
The time-varying terms ($C_1^\pm$) represent sidebands of the resonance created by the modulated coupling, which allow waves with a frequency offset $\Omega$ from the resonance frequency to couple into the resonance. We can now find the steady-state fields $b_2$ and $b_1$ (also with $a_2 = 0$) as
\begin{align*}
    b_2 = a_1 \bigg( e^{-j \beta \ell (N-1)} 
    & - \frac{C_2^0 C_1^0 }{j (\omega - \omega_0) + \gamma}  - \frac{C_2^+ C_1^0}{j (\omega - \omega_0) + \gamma}  - \frac{C_2^- C_1^0}{j (\omega - \omega_0) + \gamma} \tag{\stepcounter{equation}\theequation}  \\
    & - \frac{C_2^0 C_1^+ }{j (\omega + \Omega - \omega_0) + \gamma}  - \frac{C_2^+ C_1^+}{j (\omega + \Omega - \omega_0) + \gamma}  - \frac{C_2^- C_1^+}{j (\omega + \Omega - \omega_0) + \gamma} \\
    & - \frac{C_2^0 C_1^- }{j (\omega - \Omega - \omega_0) + \gamma}  - \frac{C_2^+ C_1^0}{j (\omega - \Omega - \omega_0) + \gamma}  - \frac{C_2^- C_1^0}{j (\omega - \Omega - \omega_0) + \gamma} \bigg) ~ ,
\end{align*}
\begin{align*}
    b_1 = a_1 \bigg( e^{-j \beta \ell (N-1)} 
    & - \frac{C_1^0 C_1^0 }{j (\omega - \omega_0) + \gamma}  - \frac{C_1^+ C_1^0}{j (\omega - \omega_0) + \gamma}  - \frac{C_1^- C_1^0}{j (\omega - \omega_0) + \gamma} \tag{\stepcounter{equation}\theequation}  \\
    & - \frac{C_1^0 C_1^+ }{j (\omega + \Omega - \omega_0) + \gamma}  - \frac{C_1^+ C_1^+}{j (\omega + \Omega - \omega_0) + \gamma}  - \frac{C_1^- C_1^+}{j (\omega + \Omega - \omega_0) + \gamma} \\
    & - \frac{C_1^0 C_1^- }{j (\omega - \Omega - \omega_0) + \gamma}  - \frac{C_1^+ C_1^0}{j (\omega - \Omega - \omega_0) + \gamma}  - \frac{C_1^- C_1^0}{j (\omega - \Omega - \omega_0) + \gamma} \bigg) ~ .
\end{align*}
From the output field $b_2$ the transmission coefficient $S_{21} = b_2 / a_1$ and reflection coefficient $S_{11} = b_1 / a_1$ can be found (with $a_2 = 0$). 
The reverse transmission coefficient $S_{12}$ can be found as $S_{12} = b_1 / a_2$ (with $a_1 = 0$). 
Since each term $C_{1,2}^\pm$ carries a $\pm \Omega$ frequency shift, we can understand the system as taking one input and splitting it into five frequencies separated by the modulation frequency $\Omega$. Presently, we wish to only consider the solution where output terms are the same frequency $\omega$ as the input. These are the terms that are measured on a typical network analyzer, while terms that are frequency offset can effectively be dismissed as noise produced by the system. We then obtain an equation for $S_{21}$ without any frequency shifted terms:
\begin{equation}
\begin{split}
    S_{21} = e^{-i k \ell (N-1)} - \frac{C_2^0 C_1^0 }{i (\omega - \omega_0) + \gamma} - \frac{C_2^- C_1^+}{i (\omega + \Omega - \omega_0) + \gamma} - \frac{C_2^+ C_1^- }{i (\omega - \Omega - \omega_0) + \gamma} ~.
\end{split}
\end{equation}
\section{Resonant decay rate}
The decay rate $\gamma$ defines the resulting modal linewidth and is present in the denominator of all the S-parameter equations for our nonreciprocal time-varying coupler.
For a lossless resonator $\gamma$ is defined as \cite{CMT}
\begin{equation}
    2 \gamma = C_1 C_1^* + C_2 C_2^*
\end{equation}
which can be extended to lossy resonators through the addition of the intrinsic decay rate $\kappa_i$. We note that for a traveling wave resonance only the $C_m$ term coupling the resonant field to the output field $b_m$ in Eqn. (\ref{S2}) is used. Our system has time-varying $C_m$, so is it necessary to slightly modify this definition to
\begin{equation}
    2 \gamma = \langle C_1 C_1^* + C_2 C_2^* + \kappa_i \rangle
\end{equation}
where $\langle$ $\rangle$ represents the time-average. This approximation is valid because $\gamma \ll \Omega$ (resolved sideband regime) and the field in the resonator decays slowly compared to the modulation. The expression for $\gamma$ can then be simplified to 
\begin{equation}
    \begin{split}
        2 \gamma & = (C_1^0)^2 + (C_1^+)^2 + (C_1^-)^2 + (C_2^0)^2 + (C_2^+)^2 + (C_2^-)^2 + \kappa_i ~.
    \end{split}
\end{equation}
\section{Circulator}
\begin{figure} 
    \begin{adjustwidth}{-.9in}{-.9in}
    \centering
    \includegraphics[width = \linewidth]{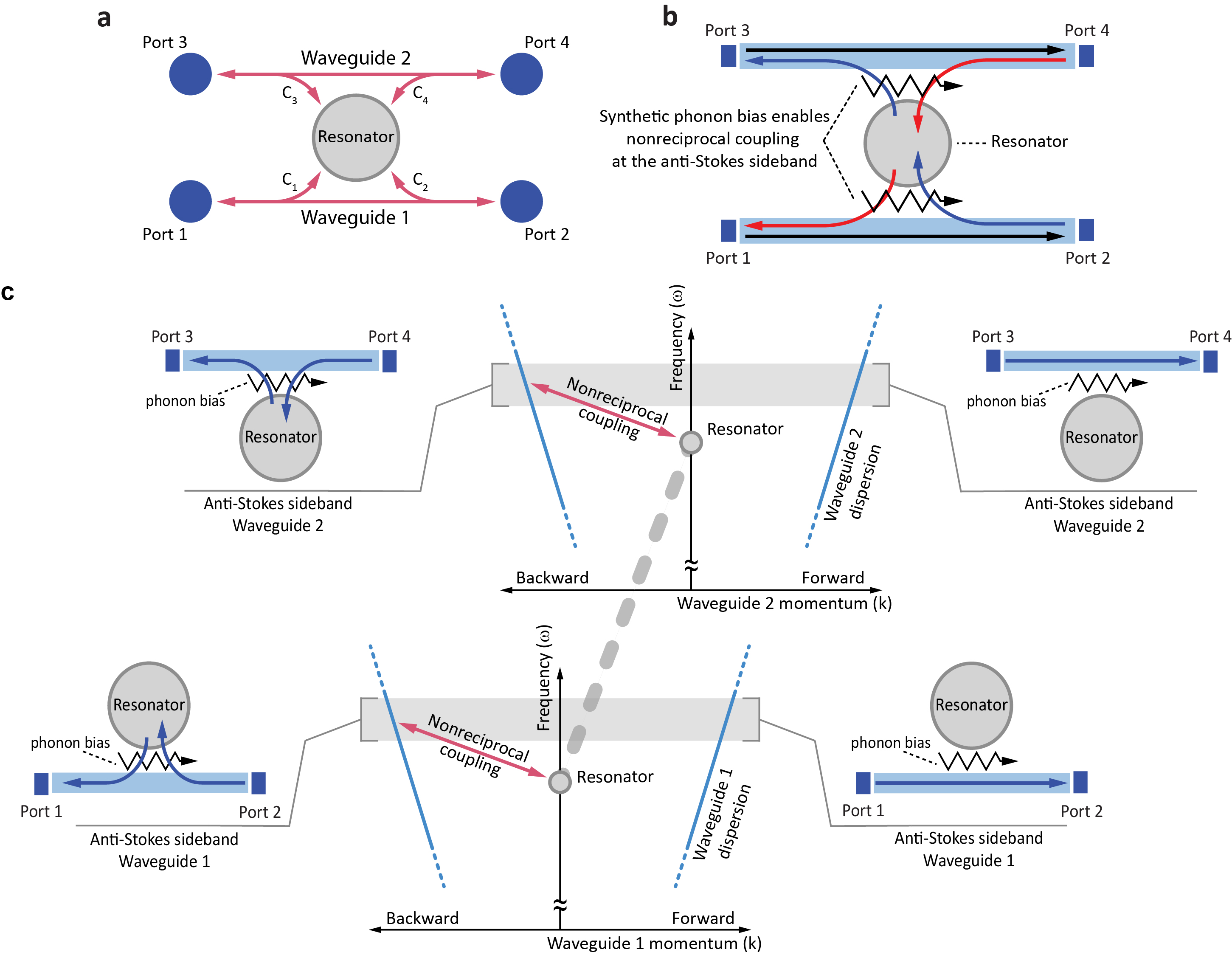}
    \end{adjustwidth}
    \begin{adjustwidth}{-.4in}{-.4in}
    \caption{\textbf{a}, Diagram of the coupling constants $C_{1-4}$ which connect the ports to the resonator.
    \textbf{b}, Schematic of the proposed circulator design. Arrows correspond to the high transmission pathways: $S_{21}$ and $S_{34}$ (black), $S_{23}$ (blue), and $S_{41}$ (red).
    \textbf{c}, Frequency-momentum diagram illustrating nonreciprocal coupling (pink arrows) at the anti-Stokes sideband, as used in the circulator design in (b). Due to the phase matching requirement, the synthetic phonon bias only couples the backward mode (port $2 \rightarrow 1$ or $4 \rightarrow 3$) in each waveguide to the resonator.}
    \label{figS2}
    \end{adjustwidth}
\end{figure}

Using two waveguides and a single resonator, it is also possible to realize a four-port circulator through synthetic phonon enabled coupling.
We introduce additional coupling constants $C_3$ and $C_4$ between the resonator and the ports of the second waveguide (ports 3 and 4) as illustrated in Fig. \ref{figS2}a.
For this device, the transmission coefficients between ports 1 and 2 (3 and 4) are the same as in the single waveguide case, and transmission through the resonator takes the form
\begin{equation} \label{s31}
    S_{31} = - \frac{C_3^0 C_1^0 }{i (\omega - \omega_0) + \gamma} - \frac{C_3^- C_1^+}{i (\omega + \Omega - \omega_0) + \gamma} - \frac{C_3^+ C_1^- }{i (\omega - \Omega - \omega_0) + \gamma} ~,
\end{equation}
since there is no direct path connecting the ports.
As in the main manuscript, we consider a device that functions at the anti-Stokes sideband frequency (Fig. \ref{figS2}b,c). If the resonator is critically coupled to the backward direction of each waveguide such that $C_1^+ C_2^- = C_3^+ C_4^- = \gamma$ and $C_1^- C_2^+ = C_3^- C_4^+ = 0$, the scattering matrix takes the form
\begin{equation}
    |S(\omega_0 + \Omega)|^2 = \begin{pmatrix}
    0 & 0 & 0 & 1 \\
    1 & 0 & 0 & 0 \\
    0 & 1 & 0 & 0 \\
    0 & 0 & 1 & 0 \end{pmatrix} ~,
\end{equation}
which is the definition of an ideal four-port circulator. 
Note that due to the additional coupling terms between the resonator and second waveguide, decay of the resonant mode increases proportionally as 
\begin{equation}\label{circDecay}
    2 \gamma = \langle C_1 C_1^* + C_2 C_2^* + C_3 C_3^* + C_4 C_4^* + \kappa_i \rangle ~.
\end{equation}
From this equation we can see that this ideal device can only be realized with a lossless resonator ($\kappa_i = 0)$ since $2 \gamma \geq C_1^+ C_2^- + C_3^+ C_4^- + \kappa_i$, although it can be approximated by making $\kappa_i$ small compared to the coupling rates. 
We implemented a four-port circulator using a similar circuit as in the main manuscript, but with an additional waveguide having three coupling sites (Fig. \ref{figS2}d). 
Synthetic phonons were applied to each trio of coupling sites with the same bias voltage to ensure symmetric coupling $C_1 = C_3$ and $C_2 = C_4$. 
The measured transmission (Fig. \ref{figS2}e) shows clear circulation behavior, with high transmission from ports $1 \rightarrow 2$, $2 \rightarrow 3$, $3 \rightarrow 4$ (not shown), and $4 \rightarrow 1$, and low transmission in the opposite direction.
Since the intrinsic loss rate in the resonator ($\kappa_i$) is not negligible in our experiment, we cannot currently achieve the critical coupling condition for circulation.
The presence of coupling at the Stokes sideband and the original resonance frequency, although small, also increases the decay rate as described by Eq. (\ref{circDecay}).
Due to these limitations, the circulator exhibits low contrast between $S_{21}$ and $S_{12}$, and the measured transmission between ports on opposite waveguides is highly lossy. 
These challenges could be addressed with a higher Q-factor resonator or by increasing the nonreciprocal coupling rate. Additionally, by increasing the number of coupling sites ($N$) and thereby making the phase matching condition more strict, coupling at the Stokes sideband and original resonance frequency can be decreased. The increased number of coupling sites would also increase the nonreciprocal coupling rate, which is proportional to $N^2$.
Despite the current challenges, this experiment provides clear evidence that circulation using synthetic phonon enabled coupling is possible. 

\begin{figure} 
    \begin{adjustwidth}{-.4in}{-.4in}
    \centering
    \includegraphics[width = \textwidth]{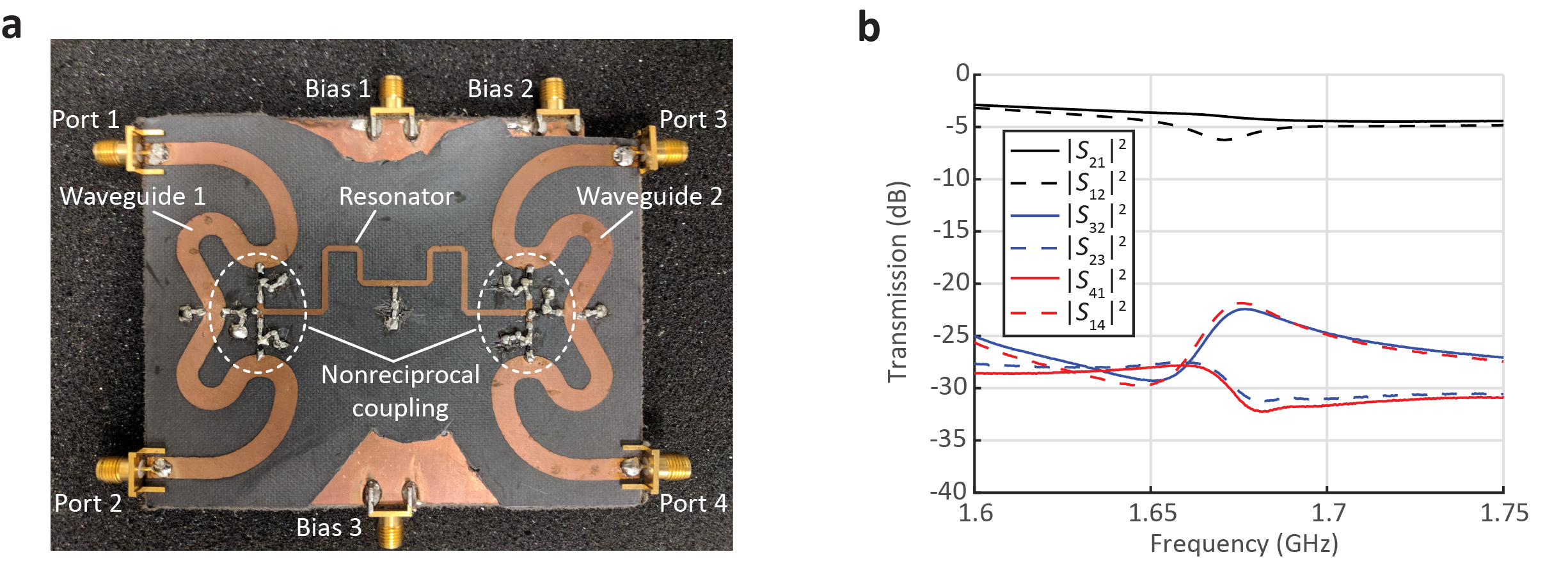}
    \caption{\textbf{a}, Photograph of the circuit implementing the circulator design in Fig. \ref{figS2}(b) with three coupling sites between each waveguide and the resonator.
    \textbf{b}, Measured power transmission for the circuit in (a), under a synthetic phonon bias which maximizes $C_1^+ C_2^-$. Colors correspond to the arrow colors in Fig. \ref{figS2}(b). Unshown transmission coefficients $S_{43}$ and $S_{34}$ are equal to $S_{21}$ and $S_{12}$ respectively, due to mirror symmetry. All other unshown transmission coefficents are reciprocal.}
    \label{figS2.2}
    \end{adjustwidth}
\end{figure}

\newpage

\end{document}